\documentclass[a4paper,twoside,12pt]{article}
\usepackage{graphicx}
\usepackage{bm}
\usepackage[left=15mm,right=15mm,top=20mm,bottom=30mm,bindingoffset=1.5cm]{geometry}

\usepackage{indentfirst}           
\usepackage{epstopdf}
\usepackage{fancyhdr}
\title{\bf Correlations in multiparticle production }  
\author{S.M.~Troshin and  N.E.~Tyurin}
\date{}
\begin{document}

\maketitle
\begin{center}
{\em 
Institute for High Energy Physics,
Protvino, Moscow Region, 142281, Russia
}
\end{center}

\vspace{5mm}

\begin{abstract}
We  discuss correlations in the hadron production in the $pp$-collision with emphasize
on the ridge-like structure origin  in the two-particle correlation function. We suggest that this structure can appear due to 
 a rotating nature of the transient state of matter generated in the intermidiate stage of proton collison. 
\end{abstract}

\newcommand{\ve}[1]{{\bm{#1}}}
\newcommand{\vegr}[1]{{\bm{#1}}}

\section*{Introduction}

Nowadays the LHC  is collecting data and providing  experimental results at the largest world
energy $\sqrt{s}=7$ $TeV$. 
Along with realization of its discovery potential,  the LHC experimental program  renewed interest to the well
known unsolved problems  providing deepened insights into those issues.
In this context the multiparticle production studies  bring us  a clue
to the mechanisms of confinement and hadronization. Confinement of a color (i.e. the fact that an isolated color object has an infinite energy in the physical vacuum)
is associated with collective, coherent interactions of quarks and gluons,
and results in formation of the asymptotic states, which are the
colorless, experimentally observable particles. 
The inelastic processes involve  large number
of particles in the final state. 

On  the other side, the experimental measurements often reveal high degrees of a coherence 
 in the relevant observables. No doubt these collective effects are very important for understanding of the nonperturbative collision dynamics.
 Such collective effects are associated, in particular, with
unitarity regulating the relative strength
of elastic and inelastic processes and connecting the amplitudes of the various multiparticle production processes.

Thus, it seems to be important now to analyze  recent experimental 
data and try to make  first conclusions on the nature of a  matter produced in
$pp$  collisions, i.e. is it  weakly interacting or it remains to be a strongly interacting one
as it was observed at RHIC in $AA$ collisions? In the latter case one can expect that proposed in \cite{intje}
mechanism related to the rotation of transient matter should be working at the LHC energies and 
therefore  the observed at RHIC phenomena should be observed in $pp$-collisons also.

\section*{Ridge-like structure in the correlation function and rotation of quark-pion liquid}
The ridge structure was observed first at RHIC in peripheral  collisons of nuclei in the two-particle 
correlation function in the near-side
jet production (cf. recent paper \cite{rhicr} and references therein). It was demonstrated that the ridge particles have
 a narrow $\Delta\phi$ correlation distribution (where $\phi$ 
is an asimuthal angle) and wide $\Delta\eta$ correlations ($\eta$ is a pseudorapidity). The ridge phenomenon was associated
with the collective effects of a medium. 

The similar structure in the two-particle correlation function was observed by
the CMS Collaboration \cite{ridgecms}. This is rather surprising result because the ridge structure was observed for the first time
in $pp$--collisions. Those collisions are commonly treated as  a kind of ``elementary'' ones under the heavy-ion studies 
and therefore often used  as the reference
process for detecting deconfined phase formation on the base of difference between $pp$- and $AA$-collisions. 
It is evident now that such approach should be revised in view of this new and  
unexpected  experimental result. This ridge structure has been observed in the small number of events in the definite
kinematical region. In addition, those events have certain limitations on their multiplicity. Despite those facts this
discovery can be compared with discovery of power-like dependence on transverse momenta of the inclusive cross-sections
which demonstrated a presence of the hard interactions (and has also been seen  in the comparatively small number of events).

The particle
production machanism  proposed in the model  \cite{intje} takes into  account  the 
geometry  of the overlap region and dynamical properties of
the transient state in hadron interaction. This picture assumes  deconfinement at the initial stage of interaction.
The transient state appears as a rotating medium of massive quarks and pions which hadronize and 
form multiparticle final state.  Essential point for this 
rotation is the non-zero impact parameter in the collision.

Indeed, the inelastic overlap function $h_{inel}(s,b)$,

\[
h_{inel}(s,b)\equiv\frac{1}{4\pi}\frac{d\sigma_{inel}}{db^2},
\]
has a peripheral impact parameter dependence at the energy $\sqrt{s}=7$ TeV due to the 
reflective scattering \cite{intja}.
Note, that unitarity equation rewritten at high energies
for the elastic amplitude $f(s,b)$ has the form
\[
\mbox{Im} f(s,b)=h_{el}(s,b)+ h_{inel}(s,b)
\]
and $h_{inel}(s,b)$ is the sum of all inelastic channel contributions.
Due to this peripherality, the mean multiplicity 
\[
 \langle n\rangle (s)=\frac{\int_0^\infty bdb  \langle n\rangle (s,b) h_{inel}(s,b)}
{\int_0^\infty bdb h_{inel}(s,b)}
\]
gets the main contribution from the collisions with non-zero impact parameters.
Thus, one can assume that the events with high multiplicity at the LHC energy $\sqrt{s}=7$ TeV  correspond
to the peripheral hadron collisions \cite{intja}. Thus, at the LHC energy $\sqrt{s}=7$ TeV there is a dynamical
selection of peripheral region in impact parameter space responsible for the inelastic processes. In the nuclear reactions
 such selection is provided by the relevant  experimental adjustments. Note, that  the ridge-like structure in the
nuclear reactions has also been observed in peripheral collisions only \cite{rhicr}.

The  geometrical picture of hadron collision at non-zero impact parameters
implies that the generated massive
virtual  quarks in overlap region  could obtain very large initial orbital angular momentum
at high energies. Due to strong interaction
between quarks this orbital angular momentum  leads to the coherent rotation
of the quark system located in the overlap region as a whole  in the
$xz$-plane (Fig. 1). This rotation is similar to the liquid rotation
where strong correlations between particles momenta exist. 
Thus, the orbital angular momentum should be realized  as a coherent rotation
of the quark-pion liquid  as a whole.
The assumed particle production mechanism at moderate transverse
momenta is an excitation of  a part of the rotating transient state of  massive constituent
quarks (interacting by pion exchanges). 
\begin{figure}[h]
\begin{center}
\resizebox{7cm}{!}{\includegraphics*{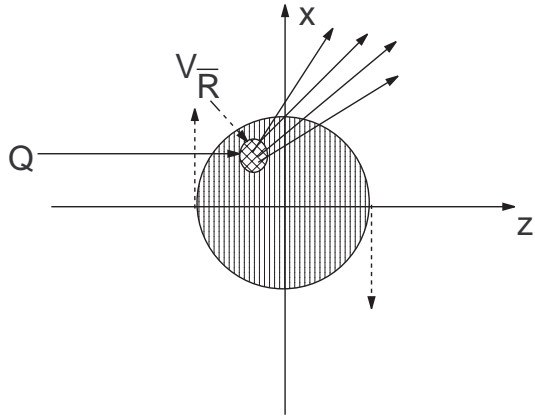}}
\caption{\small \it {Interaction of the constituent quark with rotating quark-pion liquid.}}
\end{center}
\end{figure}
Due to the fact that the transient matter is strongly interacting, the excited parts
should be located closely  to the periphery  of the rotating transient state otherwise absorption
 would not allow to quarks and pions  leave the interaction region (quenching). 

The mechanism is sensitive
 to the particular  direction of rotation and to the rotation plane orientatation. This will
lead to the narrow distribution of the two-particle correlations in $\Delta\phi$. However,
two-particle correlation could have broad distribution in polar angle ($\Delta\eta$) in the above mechanism (cf. Fig. 1).
 Quarks in the exited part of the cloud
could have different values of the two components of the momentum (with its third component
lying in the rotation  $xz$-plane) since the exited region $V_{\bar R}$ has significant extension. 

Thus,  the ridge-like structure observed  in the high multiplicity events by the CMS Collaboration 
can be an experimental manifestation of the coherent rotation of the transient matter in hadron
collisions. The narrowness of the two-particle correlation distribution in the asimuthal angle is the
distinctive feature of this mechanism. 

There should be other experimentally
observed effects of this collective effect, one of them is the directed flow $v_1$ in hadron reactions,
with fixed impact parameter discussed in \cite{intje}.  Rotation of transient matter
will affect also elliptic flow  $v_2$ and  average transverse momentum of secondary particles produced in proton-proton collisions \cite{mpla,mpla1}.
Due to  rotation, in particular, 
the following relation of the average transverse momentum with the mean multiplicity of secondaries
\[
 \langle p_T\rangle(s)=a+b\langle n\rangle (s)                                                                                            
\]
takes place.
This relation is in a good agreement with existing experimental data \cite{mpla}.

The above discussion shows that the nature of the state of matter 
revealed at the LHC in proton collisions is the same as the nature of the state revealed
 at RHIC in nuclei collisions.  The details and discussions of the other collective effects
such as elliptic flow can be found in \cite{coll}.
\section*{Conclusion}
To conclude we would like to note that the vanishing anisotropic flows and
appearance (simultaneous) of the secondary particles polarization (cf.  \cite{coll})  are
the signals of the genuine QGP formation in the mechanism based on the rotating behaviour of the transient state. 
The leading role belongs to orbital angular momentum in the above
 phase transition.
\section*{Acknowledgements}
One of the authors (S.T.) would like to thank organisers of the  15th RDMS Conference in Alushta, in particular Leonid Levchuk and Anatoly Zarubin,
for their support and warm hospitality  and Victor Kim and Igor Lokhtin for the interesting discussions.

\end{document}